\documentclass[11pt,a4paper]{article}
\usepackage[margin=24mm]{geometry}
\usepackage[T1]{fontenc}
\usepackage[utf8]{inputenc}
\usepackage{newtxtext,newtxmath}
\usepackage[nopatch=footnote]{microtype}
\usepackage{graphicx}
\usepackage{booktabs,array,tabularx,longtable}
\usepackage[numbers,sort&compress]{natbib}
\usepackage{caption}
\usepackage{xurl}
\usepackage[hidelinks]{hyperref}
\hypersetup{pdftitle={Local error correction prolongs time-crystalline memory}}
\usepackage{amsmath}
\usepackage{setspace}
\usepackage{authblk}
\graphicspath{{figures/}}

\newcommand{\nd}{N_{\mathrm{data}}}
\newcommand{\na}{N_{\mathrm{ancilla}}}
\newcommand{\Cpi}{C_{\pi}}
\newcommand{\Gc}{G_{\mathrm c}}

\title{Local error correction prolongs time-crystalline memory}
\author{Benedikt Fauseweh}
\affil{Department of Physics, TU Dortmund University, 44227, Dortmund, Germany}
\date{\today}

\begin{document}
\maketitle

\begin{abstract}
Discrete time crystals are non-equilibrium phases of matter with spatiotemporal order and a response that breaks the time-translation symmetry of a periodic external drive. Interactions, spatial disorder and correlated random driving can in theory stabilize this order. In experiments, however, coupling to the environment limits the time crystal lifetime. Here we use stochastic local bit-flip error correction to prolong the time-crystalline response of a driven, spatially disordered two-dimensional Ising model on a superconducting quantum processor, using a repetition code with distances three and five on a honeycomb lattice with up to 68 physical qubits. Even though ancilla qubits, midcircuit syndrome measurements and decoding increase the required quantum and classical resources, the stochastic drive increases the memory decay time by a factor of up to two. At late times we observe that recovery stabilizes spatial order across the lattice. The spin-configuration distributions explain this separation, as recovery suppresses disagreeing bonds while broadening occupies completely reversed states. Syndrome measurements without correction accelerate the memory decay, demonstrating the importance of conditional correction. Our results show that local bit-flip error correction can preserve time-crystalline memory over more drive cycles than the bare dynamics.
\end{abstract}

\section*{Introduction}

Quantum simulators use controllable quantum systems to study the behavior of strongly interacting quantum systems~\cite{feynman_simulating_1982,cirac_goals_2012}. Digital processors realize this evolution through sequences of quantum gates~\cite{lloyd_universal_1996}, offering a flexible route to studying dynamical regimes that challenge classical calculations~\cite{daley_practical_2022,Kim2023,doi:10.1126/sciadv.adk4321,barron2026observableestimationabsenceclassical}. Fundamental phenomena, such as the thermalization of quantum systems and non-equilibrium induced long range order, unfold over long time scales and require large system sizes to avoid finite-size effects~\cite{eisert_quantum_2015,DAlessio03052016}. On a noisy processor, gate errors and coupling to the environment progressively alter the evolving state~\cite{preskill_quantum_2018,fauseweh_quantum_2024}. Extending the interval over which the physical signal survives is therefore a central challenge for quantum simulation.

Quantum error correction offers a route to extending this time window. Repeated parity checks can identify flip errors without measuring the encoded orientation, as demonstrated in repetition-code memories~\cite{google_repetition_2021}. Recent experiments have demonstrated below threshold quantum memories and performed protected operations on logical information~\cite{google_below_threshold_2025,putterman_bosonic_2025,sales_rodriguez_magic_2025,bluvstein_fault_tolerant_2026}. These results motivate using measurement and conditional correction within a quantum simulation to extend coherent many-body dynamics beyond the normal decoherence time of the device. 

A prime candidate to test this hypothesis are time crystals. Time crystals are non-equilibrium phases of matter that break time-translational symmetry, causing a system to oscillate indefinitely in a stable, repeating pattern. Unlike conventional crystals whose atoms repeat periodically in space, time crystals realize order in both, space and time. Because equilibrium systems cannot spontaneously break continuous time-translation symmetry~\cite{watanabe_absence_2015}, realizing this order requires driving the system out of equilibrium~\cite{shapere_classical_2012,wilczek_quantum_2012,sacha_time_2018,else_discrete_2020}. In a driven spin system, interacting spins can keep alternating between opposite orientations even when the pulses intended to reverse them are imperfect. In a discrete time crystal, this response remains locked to multiples of the drive period over a range of pulse parameters, leading to subharmonic correlations with respect to the driving frequency~\cite{else_floquet_2016,khemani_phase_2016,yao_discrete_2017}. Subharmonic spin responses have been observed in trapped ions, spins in diamond and superconducting processors~\cite{zhang_observation_2017,choi_observation_2017,randall_many-bodylocalized_2021,kyprianidis_observation_2021,mi_time-crystalline_2022,zhang_digital_2022,frey_realization_2022}. 
Importantly it was recently shown, that suitably correlated random drives support temporal order at fixed observation times alongside irregular motion between them~\cite{zhao_temporal_2023,Moon2025}.

For the oscillation to persist, the spins must retain order as the drive supplies energy. In a closed, one-dimensional disordered spin system, localization can suppress heating and allow the subharmonic response to persist~\cite{else_floquet_2016,khemani_phase_2016}, while for higher dimensions fluctuations are predicted to destabilize many-body localization at sufficiently large scales~\cite{de_roeck_stability_2017}. Another scenario which leads to exponentially long lifetimes are high frequency drives which support a long prethermal interval before appreciable heating~\cite{else_prethermal_2017}. However in open systems, such as noisy quantum processors, the coupling to the environment and imperfect operations further limit that interval~\cite{lazarides_fate_2017,ippoliti_many-body_2021}. Extending it requires countering errors while retaining the spatial spin pattern and its oscillating response.

Recent theoretical works predict that local measurements and feedback can stabilize temporal order or prolong its lifetime under noise~\cite{PhysRevB.104.L180304,mcginley_absolutely_2022,camacho_prolonging_2024}. Here we implement local bit-flip error correction during the evolution of a driven, spatially disordered Ising model on a superconducting processor. Ancillas on a heavy-hex lattice mediate Ising interactions and measure bond parities on honeycomb patches with up to 68 physical qubits. Conditional data gates implement a stochastic repetition-code recovery on blocks of three or five sites, which introduces temporal disorder into the recovery schedule. We investigate the effects of varying activation probability and system size upon time crystal memory, spatial correlation functions and measurement distributions. At high activation probability, recovery extends the fitted temporal memory lifetime by up to a factor of 2.01 and sustains spatial correlations across each patch. A control with parity measurements alone shows faster memory decay, demonstrating the benefit of conditional correction. The measured configurations also show that spatial alignment persists more strongly than temporal phase memory. Our approach establishes a net lifetime benefit of local recovery for a non-equilibrium quantum simulation.

\section*{Results}

We investigate a time crystal model on honeycomb patches with open boundaries. The smaller patch contains two adjacent hexagons, 10 data sites and 11 data bonds. The larger patch contains a $3\times3$ array of hexagons, 30 data sites and 38 bonds. Each bond consists of an ancilla qubit between its endpoints on the physical processor. The two experiments therefore use 21 and 68 physical qubits, respectively. 

\begin{figure}[p]
\centering
\includegraphics[width=\linewidth,height=0.73\textheight,keepaspectratio]{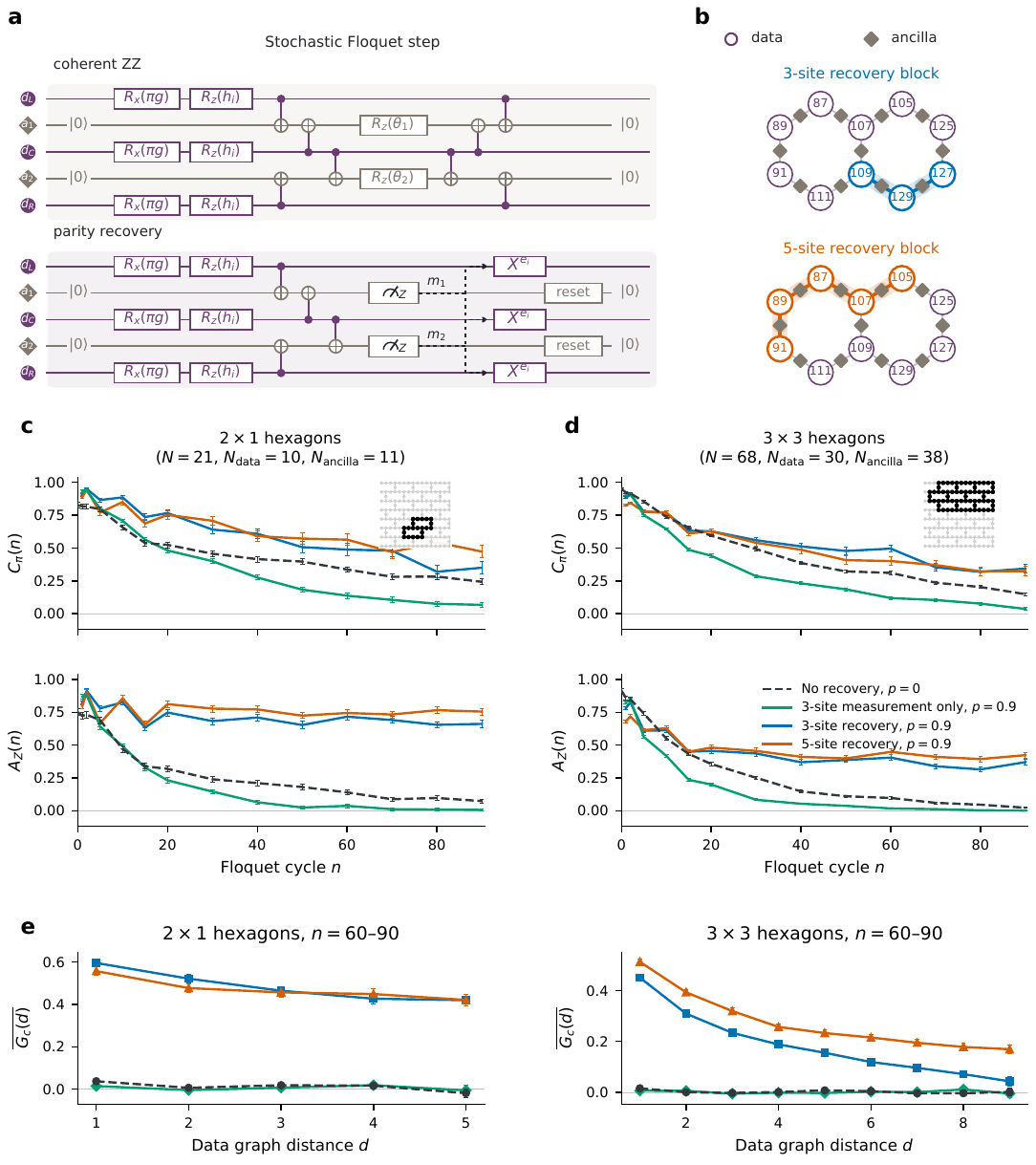}
\caption{\textbf{Local bit-flip error correction extends temporal and spatial order.}
\textbf{a}, The stochastic drive step has two alternatives for a candidate recovery block. With probability $1-p$, entangling gates and local rotation gates implement coherent ZZ interactions. With probability $p$, parity measurements and conditional data flips implement recovery, followed by ancilla reset. 
\textbf{b}, Physical embeddings of example three-site and five-site recovery blocks on two adjacent hexagons. Circles are data qubits and diamonds are ancillas.
\textbf{c,d}, Initial state memory and mean alignment $A_Z$ for the two patches. Insets mark the selected hardware qubits in black.
\textbf{e}, Connected correlations versus distance on the data graph, averaged equally over the sampled cycles $60,70,80,90$. Bare evolution is compared with three-site recovery, five-site recovery and three-site measurements alone. Recovery blocks are activated with probability $p=0.9$. All runs use $g=0.97$, quenched disorder and 256 shots per circuit. Bars show one sampling standard error after readout correction, conditional on the stored calibration.}
\label{fig:overview}
\end{figure}

Our unitary drive is a Floquet Ising model which combines a nearly global spin inversion with Ising interactions and longitudinal fields. The drive is applied in discrete steps, each of which is implemented by a sequence of quantum gates on the processor, see Fig.~\ref{fig:overview}a.
\begin{equation}
 U_F =
 \exp\!\left(-\frac{i}{4}\sum_{\langle ij\rangle}J_{ij}Z_iZ_j\right)
 \exp\!\left(-\frac{i}{2}\sum_i h_i Z_i\right)
 \exp\!\left(-\frac{i\pi g}{2}\sum_i X_i\right).
 \label{eq:drive}
\end{equation}
For the disordered system the couplings and fields are uniformly distributed according to $J_{ij}\in[-3\pi/2,-\pi/2]$ and $h_i\in[-\pi,\pi]$. The rotation angles in the circuit are $R_{ZZ}(J_{ij}/2)$ and $R_Z(h_i)$. Equation~\eqref{eq:drive} extends the disordered Floquet Ising model used in studies of time-crystalline eigenstate order to the honeycomb geometry~\cite{khemani_phase_2016,ippoliti_many-body_2021,mi_time-crystalline_2022}. The kick parameter $g$ is crucial for the subharmonic response. For $g=1$, the drive inverts all spins perfectly, however with strong disorder also $g<1$ leads to a stable time crystal in the one-dimensional case, with $g_c \approx 0.84$, below which the subharmonic response is lost. 

The unitary step repeats every period $T$, while recovery varies between kicks through rotating block coverings and stochastic activation. At kick $n$, the covering $\mathcal B_n$ specifies disjoint eligible blocks, see Fig.~\ref{fig:overview}b. Each block $B$ is activated independently with probability $p\in[0,1]$. An activated block of $k$ data qubits measures $k-1$ bond parities on a connected tree. The outcomes form a syndrome, from which a minimum-weight decoder selects the compatible pattern with the fewest spin flips. Conditional $X$ gates apply this correction. We refer to this sequence of syndrome acquisition, decoding and conditional correction as recovery. Note that this scheme is not a full error correction code, as it does not protect against phase errors. In the perfect case, i.e. $g=1$, single phase errors do not alter the ideal subharmonic response, which is a global flip between the two orientations. The recovery therefore aims to correct bit flips to maintain the subharmonic response. 

Under ideal conditions, the combined evolution is
\begin{equation}
 \rho_{n+1}
 =
 \left[
 \prod_{B\in\mathcal B_n}
 \bigl((1-p)\mathcal I+p\mathcal R_B\bigr)
 \right]
 \left(U_F\rho_nU_F^\dagger\right),
 \label{eq:recovered_drive}
\end{equation}
where $\rho_n$ is the data-qubit density matrix averaged over syndrome outcomes and $\mathcal I$ is the identity channel. In each hardware circuit, we fix the activation mask before compilation. The block channel is by
$\mathcal R_B(\rho)=\sum_{\mathbf s}
F_{B,\mathbf s}\Pi_{B,\mathbf s}\rho
\Pi_{B,\mathbf s}F_{B,\mathbf s}^{\dagger}$.
Here $\Pi_{B,\mathbf s}$ projects onto the joint parity outcome $\mathbf s$, and $F_{B,\mathbf s}$ applies the spin flips selected by the decoder. Under ideal operations, an activated block corrects any pattern of up to one bit flip for $k=3$ and up to two for $k=5$, relative to either reference orientation. For each case $k$, $p$ and system size we averaged over $256$ hardware shots. The repetition-code recovery treats a spin configuration and its global complement equally, since both have the same bond parities. Consequently, the rule also preserves the two orientations needed for a subharmonic response. 

A useful hardware property is that we can employ the ancilla qubits for both interaction and recovery circuits. Two controlled-NOT gates project $Z_iZ_j$ into a bond ancilla. An ancilla $Z$ rotation followed by uncomputation realizes the coherent Ising interaction. For an activated parity check, the circuit measures the ancilla for the decoder. No further rotation is required, as it becomes a global phase.

In the hardware experiments we investigate the $|0\rangle^{\otimes\nd}$ state as initial state. We further analyze other $Z_i$ initial states through classical simulations, see Supplementary Information S5. 

We quantify temporal memory by
\begin{equation}
 \Cpi(n)=\frac{(-1)^n}{\nd}\sum_i z_i(0)\langle Z_i(n)\rangle,
 \label{eq:memory}
\end{equation}
where $z_i(0)$ is the initial value of $Z_i$. A perfect subharmonic response has $\Cpi=1$, while quantum fluctuations and decoherence typically reduce this observable. We quantify spatial alignment by the mean correlation between distinct data-spin pairs,
\begin{equation}
 A_Z(n)
 =\frac{2}{\nd(\nd-1)}\sum_{i<j}\langle Z_i(n)Z_j(n)\rangle
 =\frac{\nd\langle M_z^2(n)\rangle-1}{\nd-1},
 \label{eq:alignment}
\end{equation}
where $M_z=\nd^{-1}\sum_i Z_i$. Here $A_Z=0$ for independent unpolarized spins and $A_Z=1$ for perfect alignment. 

Figure~\ref{fig:overview}c,d compares the bare drive with three-site and five-site recovery at $p=0.9$. Both recovery schemes increase temporal memory. Averaged equally over the sampled cycles $60,70,80,90$, the smaller patch has $\overline{\Cpi}=0.29\pm0.01$ for the bare drive, $0.41\pm0.02$ for three-site recovery and $0.51\pm0.02$ for five-site recovery. On the larger patch, the corresponding values are $0.23\pm0.01$, $0.38\pm0.02$ and $0.36\pm0.02$. Fitting an exponential $A\exp(-n/\tau)$ across the measured positive depths yields $\tau=66\pm2$, $88\pm6$ and $133\pm12$ cycles for the smaller patch. The larger patch gives $49\pm1$, $85\pm4$ and $82\pm4$ cycles. The lifetimes $\tau$ are measured in drive cycles and quantify the persistence of the subharmonic response.

We can also estimate the physical time over which the measured response persists using the gate, midcircuit measurement and reset times. Multiplying the cycle-domain decay scales by the estimated mean cycle durations gives approximately $39.7$, $420$ and $623\,\mu\mathrm{s}$ for the smaller patch and $35.4$, $416$ and $399\,\mu\mathrm{s}$ for the larger patch, for bare dynamics, three-site recovery and five-site recovery respectively. These are conversions under the optimized scheduling model and not directly measured device lifetimes. The cycle timing and lifetime conversion are detailed in Supplementary Section S6.

Spatial alignment shows an even more pronounced effect, as it stabilizes at finite value above the bare response. In the same window, the smaller patch has $\overline{A_Z}=0.099\pm0.008$ for the bare drive and $0.75\pm0.01$ with five-site recovery. The larger patch increases from the bare $0.056\pm0.003$ to $0.41\pm0.01$ with five-site recovery. This observation shows that local recovery has a stronger effect on spatial alignment than on temporal memory, as the subharmonic response requires both local alignment and a global phase relationship between the two orientations.

Repeated measurements without corrections can themselves influence the lifetime of temporal order. Theoretical calculations predict that measuring the total magnetization can prolong a many-body localized discrete time crystal response in a spin chain coupled to a thermal bath~\cite{cenedese_thermodynamics_2025}. To separate parity acquisition from active correction in our local scheme, we run a measurement control with the same three-site block selection, interaction circuits and ancilla resets but without conditional data corrections. Figure~\ref{fig:overview} shows that this control has lower memory than the bare drive. Spatial correlations remain small as well and we therefore conlcude that the enhanced order is due to the full conditional recovery.

To further understand the spatial structure, we investigate the connected correlation function, which spatially resolves fluctuations beyond the individual spin polarizations,
\begin{equation}
 \Gc(d,n)=\frac{1}{|\mathcal P_d|}
 \sum_{(i,j)\in\mathcal P_d}
 \left[\langle Z_iZ_j\rangle_n-\langle Z_i\rangle_n\langle Z_j\rangle_n\right],
 \label{eq:connected}
\end{equation}
in Fig.~\ref{fig:overview}e. Here the set $\mathcal P_d$ contains site pairs at graph distance $d$ on the data lattice. Correlations are averaged over the sampled cycles 60 to 90. Five-site recovery gives $\overline{\Gc}(5)=0.421\pm0.027$ across the diameter of the smaller patch. Across the larger patch, $\overline{\Gc}(9)=0.17\pm0.02$. The bare values are consistent with zero at both diameters. The larger patch shows correlated fluctuations over distances exceeding a recovery block. While the stochastic choice of the recovery block can lead to a larger region of correlated fluctuations, it is important to note that also a mixture of the two aligned orientations contributes to $\Gc$. To resolve this we analyze the distribution of measured configurations in the $Z$ basis, see below.

\begin{figure}[p]
\centering
\includegraphics[width=\linewidth]{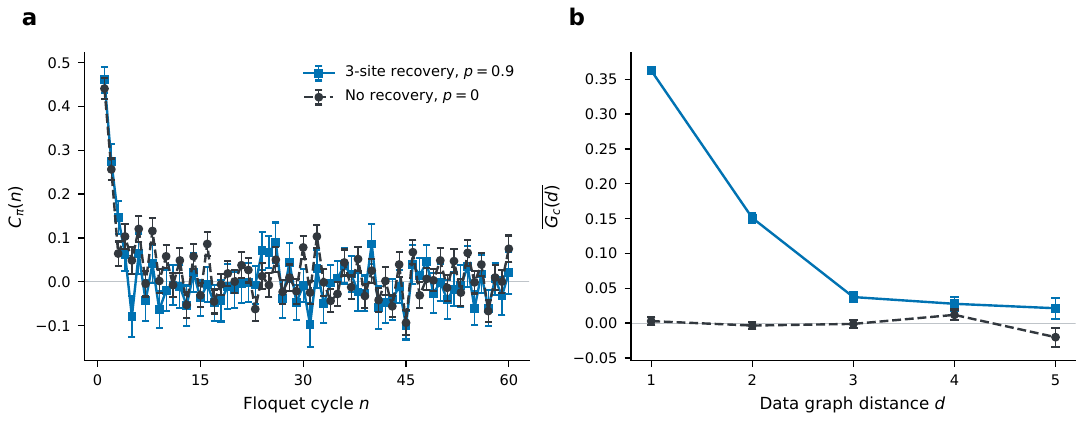}
\caption{\textbf{Small pulse angle memory decay.}
\textbf{a}, Initial-state memory on two hexagons with ten data qubits and eleven ancillas at $g=0.67$. Quenched interactions and fields use the same disorder intervals as Fig.~\ref{fig:overview}. Bare evolution is compared with full three-site recovery at $p=0.9$. Every circuit starts from $|0000000000\rangle$, with 128 shots per condition at each depth from 1 to 60.
\textbf{b}, Connected correlation averaged equally over cycles 40 to 60 after averaging pairs at each graph distance. Recovery maintains short-distance correlations while subharmonic memory is lost. Bars indicate one sampling standard error.}
\label{fig:uniform}
\end{figure}

We next examine whether frequent recovery sustains the alternating response at a substantially different pulse angle. On the ten-site patch, we set $g=0.67$ while retaining quenched interactions and fields from the same disorder intervals used in Fig.~\ref{fig:overview}. Every circuit starts from $|0\rangle^{\otimes 10}$. Bare evolution and three-site recovery at $p=0.9$ both rapidly lose temporal memory, see Fig.~\ref{fig:uniform}. Averaged over cycles 40 to 60, they both result in $\overline{\Cpi} \approx 0$. Recovery nevertheless maintains a nearest-neighbor connected correlation of $0.36\pm0.01$. Across the diameter of the patch, the correlation decays to zero. We conclude that strong local alignment can persist after subharmonic memory is lost, even with spatial disorder present. The local recovery rule alone is insufficient to sustain the alternating response or long range order at this drive setting. In a generic ergodic Floquet system, heating erases local memory of the initial state~\cite{lazarides_equilibrium_2014}. The rapid loss of $\Cpi$ is consistent with approaching this regime as the pulse moves away from inversion.

We further investigate the $g=0.97$ case in Fig.~\ref{fig:distributions}. The measured configurations in the $Z$ basis are used to compute the number of disagreeing bonds $W$ and the Hamming distance from the target word $K$. The distributions of these quantities are shown for the larger patch, normalized at a fixed depth and averaged over the sampled cycles 60, 70, 80 and 90.

Increasing the block activation probability shifts the bond distribution towards smaller $W$. Its mean falls from $17.7$ for the bare drive to $8.0$ and $7.2$ for three-site and five-site recovery at $p=0.9$. The width changes more modestly, from a standard deviation of $3.4$ to $3.8$ and $4.0$ respectively. Recovery therefore reduces disagreement while fluctuations remain of the same size for the domain boundaries.

The target distance behaves differently. Its mean falls from $11.6$ to $9.4$ and $9.7$, while its standard deviation grows from $2.9$ to $7.1$ and $8.2$. An increasingly ordered configuration can have either global orientation. In particular, $W=0$ is compatible with both $K=0$ and $K=\nd$. Broad target distance distributions are thus consistent with extended regions of reversed spins, which in particular can come from a proliferation of phase errors followed by the $g$ pulse leading to extended $X$ errors which can not be corrected by the repetition code. 

The intermediate value $p=0.5$ lies exactly between the bare drive and the $p=0.9$ case. The distributions therefore show that recovery reduces the number of disagreeing bonds while broadening the target distance distribution. The domain-wall fraction $D=\langle W\rangle/\na$ shows the dependence on block activation probability, see Fig.~\ref{fig:walls}). The mean wall fraction decreases as $p$ is increased and Five-site recovery produces the lower wall fraction compared to three-site recovery.

\begin{figure}[p]
\centering
\includegraphics[width=\linewidth,height=0.73\textheight,keepaspectratio]{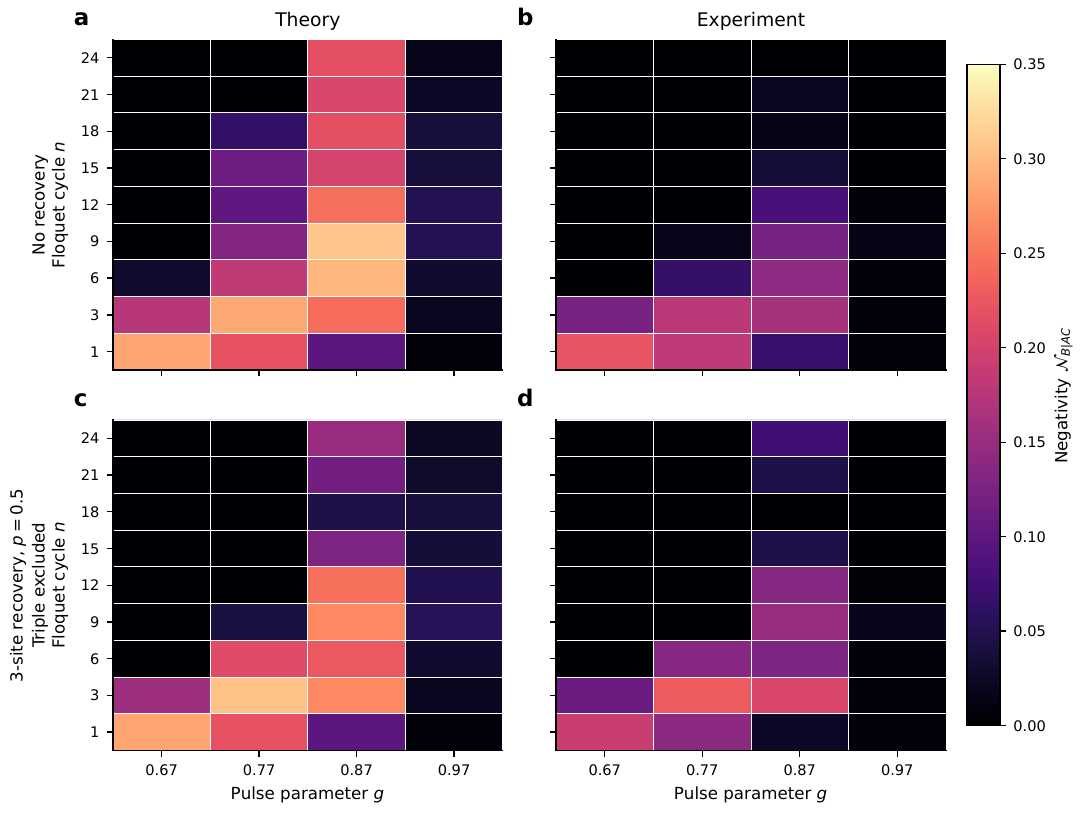}
\caption{\textbf{Local entanglement during bare and recovered dynamics.}
Negativity across $91\,|\,(89,111)$ for a fixed data-qubit triple on the two-hexagon patch.
\textbf{a,b}, Bare dynamics.
\textbf{c,d}, Three-site recovery at $p=0.5$.
\textbf{a,c}, Exact ten-data-qubit theory with ideal gates and recovery.
\textbf{b,d}, Density-matrix estimates from 27 Pauli settings with 768 shots each. All conditions start from $|0\rangle^{\otimes10}$ and share the same quenched disorder. Recovery uses the same qubit activations in theory and experiment. Colors show fitted negativity. 
}
\label{fig:negativity}
\end{figure}

Finally, we examine the quantum correlations accompanying this dynamics. Figure~\ref{fig:negativity} compares the negativity of a fixed three-data-qubit reduced state in theory and experiment~\cite{vidal_computable_2002}. For the reduced state $\rho_{ABC}$, with middle qubit $B$ and its two neighbors $A$ and $C$, the negativity is
\begin{equation}
 \mathcal N_{B|AC}(\rho_{ABC})
 =\frac{\|\rho_{ABC}^{T_B}\|_1-1}{2},
 \label{eq:negativity}
\end{equation}
where $T_B$ denotes partial transposition on $B$ and $\|\cdot\|_1$ is the trace norm. We obtain $\rho_{ABC}$ through quantum state tomography. A positive negativity certifies entanglement across the chosen partition even when the reduced state is mixed. Varying $g$ over $0.97$, $0.87$, $0.77$ and $0.67$ probes how quantum correlations develop away from the near-inversion regime. Their subsequent redistribution into the rest of the system can reduce local negativity during ergodic evolution~\cite{iemini_signatures_2016}. Together with temporal memory, its dependence on $g$ and $n$ provides a finite-time probe of the crossover from time-crystalline dynamics towards the ergodic regime.

At $g=0.67$ and $0.77$, three-site recovery at $p=0.5$ reduces the early negativity. This is consistent with the removal of coherence between parity sectors by syndrome measurement and averaging over its outcomes. At several later sampled depths, recovery gives a larger reconstructed negativity than the bare drive. Theory also shows late local entanglement with recovery, most clearly at $g=0.87$, where the bare drive has lost its negativity. This qualitative agreement indicates that recovery can change both the generation and the retention of local quantum correlations and increase the total time in which entanglement can be observed in experiment. 

\section*{Discussion}

Our results demonstrate that local recovery can enhance the lifetime of interacting quantum dynamics despite the additional classical and quantum resources required by the implementation. The combination of drive and recovery rate determine how long the discrete time crystal signal survives. At small pulse angles, temporal memory decays despite spatial disorder and frequent local correction, while short-distance alignment remains. Localization due to disorder can suppress drive-induced heating, while recovery removes spin defects during noisy evolution~\cite{ippoliti_many-body_2021,camacho_prolonging_2024}. 

Recovery also changes the structure of the residual spin defects. The broad target distance distributions and enhanced spatial correlations are consistent with extended regions of reversed spins. Reversing a connected region changes many spins while creating disagreeing bonds only at its boundary. A recovery block covering exactly that region therefore receives no information about the error. Identifying the underlying microscopic faults requires further information. Correlated hardware faults, accumulated and undetected flip errors and domain growth induced by incorrect recovery can lead to such reversed regions. Correlated error bursts have been resolved in superconducting processors through time-resolved error diagnostics~\cite{google_repetition_2021,mcewen_bursts_2022}. Works on one-dimensional cellular-automaton decoders show that local voting rules can prolong memory while continuing noise still permits error domains to grow~\cite{lang_strictly_2018,guedes_quantum_2024}. The ability to remove extended domains is therefore an important benchmark for scaling local recovery.

A separate issue is the protection of the phase. Every $Z_i$ error commutes with the measured $ZZ$ checks. Larger blocks and repeated checks therefore supply no direct phase error syndrome. Protecting coherent superpositions as well requires suppressing or correcting phase errors, e.g., through an encoding with compatible checks for both error sectors~\cite{google_below_threshold_2025}. At imperfect pulse angles, phase errors can still affect later spin populations, so their influence on the observed memory must be assessed.

Our work highlights the broader opportunity to design quantum simulators around the information that defines the physical behavior of interest. Discrete time crystals supply that information through spatio-temporal correlations, and local in-circuit measurements allow for its active stabilization. The recovery channel also provides a programmable irreversible component of the dynamics. Changing its spatial rule and rate opens a route to studying how driven order emerges, how domains relax and how temporal memory is lost in a controlled open many-body system~\cite{zaletel_colloquium_2023}. Our measurements of the negativity reveal the role of local entanglement. Extending that measurement to larger regions would help resolve how recovery changes the redistribution of entanglement near the loss of temporal memory.

\section*{Methods}

\subsection*{Processor and compilation}

In our experiments we used the 156-qubit \texttt{ibm\_kingston} Heron processor~\cite{AbuGhanem2025}. Bond ancillas mediate the honeycomb data graph within its heavy-hex connectivity. Circuits were compiled with Qiskit, with fixed placement, routing disabled and otherwise full optimization level. The gate basis is $R_Z$, $\sqrt X$, $X$ and controlled $Z$. Physical qubit assignments, couplers and calibrations are given in Supplementary Section S1.

\subsection*{Recovery implementation}

Ancillas measure bond parities and a syndrome lookup table selects conditional data $X$ gates. Ancillas are then reset and classical syndrome registers reused. Coherent interactions use four controlled $Z$ gates per bond, reduced to two when the same bond is measured. Implementation and recovery schedules are given in Supplementary Section S2. In the circuits, all commuting parity computations are arranged in three disjoint CZ matchings, followed by reverse computation on unchecked bonds.

\subsection*{Measurements and matching}

\begin{figure}[p]
\centering
\includegraphics[width=\linewidth,height=0.70\textheight,keepaspectratio]{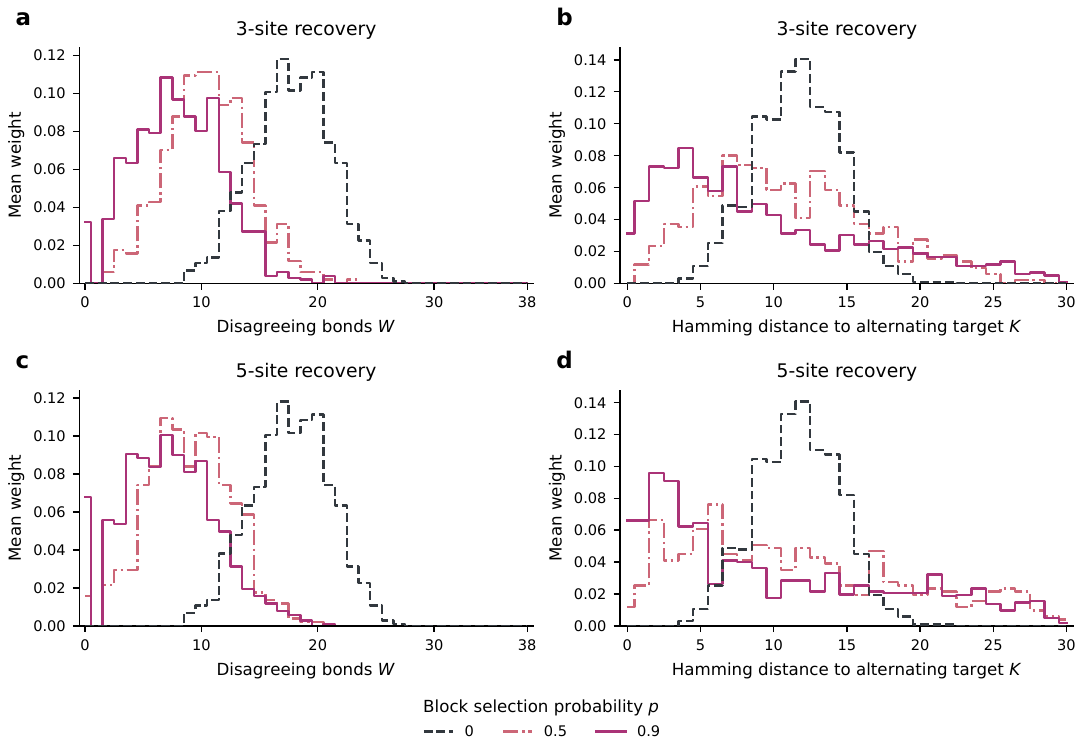}
\caption{\textbf{Probability distributions for Hamming distance and bonds.}
Mitigated distributions on the $3\times3$ patch, normalized at each depth and averaged equally over the sampled cycles $60,70,80,90$. The acquired conditions are the shared bare reference $p=0$ and recovery at $p=0.5$ and $p=0.9$. The $p=0$ and $p=0.9$ curves use 256 shots, while the $p=0.5$ curve uses 128 shots. 
\textbf{a,b}, Three-site recovery.
\textbf{c,d}, Five-site recovery.
\textbf{a,c}, Number $W$ of disagreeing data bonds.
\textbf{b,d}, Hamming distance $K$ from the target word. The two globally aligned orientations both have $W=0$ but occupy opposite ends of the $K$ axis.}
\label{fig:distributions}
\end{figure}

\begin{figure}[p]
\centering
\includegraphics[width=\linewidth]{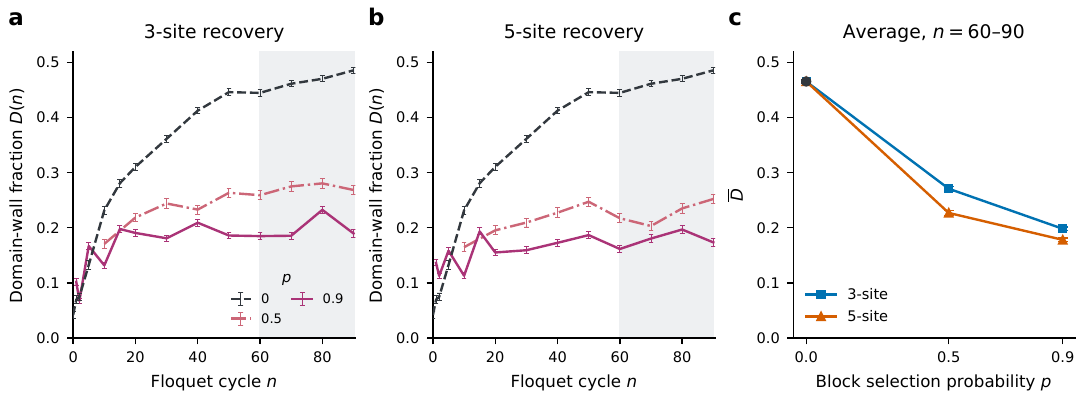}
\caption{\textbf{Spatial order dependence on block activation probability.}
\textbf{a,b}, Domain wall fraction $D(n)$ on the $3\times3$ patch for three-site and five-site recovery at $p=0.5$ and $p=0.9$, with the shared bare reference $p=0$. The shaded region marks cycles 60 to 90.
\textbf{c}, Equal-depth averages over the sampled cycles $60,70,80,90$. Five-site recovery gives a lower wall fraction at both acquired nonzero probabilities. The $p=0.5$ curves use 128 shots per circuit, all other curves use 256 shots. The black point is the bare reference.}
\label{fig:walls}
\end{figure}

The data in Figs.~\ref{fig:overview}, \ref{fig:distributions} and \ref{fig:walls} uses one fixed disorder realization per geometry and depths $0$, $1$, $2$, $5$, $10$, $15$, $20$, $30$, $40$, $50$, $60$, $70$, $80$ and $90$ on both patches. Each circuit has 256 shots. Figures~\ref{fig:distributions} and \ref{fig:walls} additionally include a separate dataset of three-site and five-site recovery at $p=0.5$ on the larger patch, with 128 shots at each depth $10,20,30,40,50,60,70,80,90$. Its embedding, disorder and reference state are the same. Hardware parameters are detailed in Supplementary Section S3.

\subsection*{Statistical analysis and timing}

Observables such as $\Cpi$ and $A_Z$ are evaluated from measurements after propagation to cycle $n$. The configuration distributions use the matrix-free measurement mitigation (M3) method~\cite{nation_scalable_2021} to mitigate readout errors. Lifetime fits minimize unweighted squared residuals of $A\exp(-n/\tau)$ over all measured depths, with positive $A$ and $\tau$ and no offset. Readout models, estimators and fit diagnostics are provided in Supplementary Section S3.

\subsection*{Tomography and readout mitigation}

Figure~\ref{fig:negativity} probes the data qubits $(89,91,111)$, with partition $91\,|\,(89,111)$. We measure all 27 product-Pauli bases with 768 shots each at $g=0.67,0.77,0.87,0.97$ and $n=1,3,6,9,12,15,18,21,24$. A convex multinomial maximum-likelihood fit reconstructs a positive, trace-one density matrix. Eight calibration circuits prepare the $2^3$ computational-basis words of the three qubits, with 768 shots per word, to estimate its full correlated assignment matrix for readout error mitigation. 

\subsection*{AI Methodology}

The author formulated the research questions and developed the research approach without AI assistance. Software and quantum circuits were implemented using ChatGPT-6 Astra and were reviewed and checked by the author. Claude Opus 5.5 was additionally used to check the generated code for bugs.
The author designed the manuscript and used ChatGPT-6 Astra to refine wording and sentence structure. Figures were prepared with assistance from ChatGPT-6 Astra. ChatGPT-6 was also used to support literature research.

\section*{Data and code availability}

The data and code that support the findings of this study are available at \url{https://doi.org/10.5281/zenodo.23338025}.

\section*{Acknowledgements}
We thank Kevin Lively and Gonzalo Camacho for helpful discussions. We acknowledge the use of IBM Quantum services for this work. The views expressed are those of the authors, and do not reflect the official policy or position of IBM or the IBM Quantum team.
This work was funded by the Deutsche Forschungsgemeinschaft (DFG, German Research Foundation)-Project No. FA 1884/5-1.

\clearpage

\clearpage

\clearpage

\clearpage

\clearpage
\end{document}